\documentclass[]{aa}
\usepackage{times}
\usepackage{graphics,amssymb,amsmath}
\usepackage{aabib99}

\begin{document}

   \thesaurus{13 (13.07.1); % GRB's
              12 (12.03.3); % Cosmology, observations  
              09 (09.04.1)} % ISM, dust, reddening

   \title{VLT identification of the optical afterglow of 
          the gamma-ray burst \object{GRB~000131} at $z=4.50$%
          \thanks{Based on observations collected at the 
                  European Southern Observatory, La Silla and Paranal,
                  Chile (ESO Programmes 64.H-0573, 64.H-0580, 64.O-0187, 
		  and 64.H-0313)}}

   \author{M.I. Andersen \inst{1}
      \and J. Hjorth \inst{2}
      \and H. Pedersen \inst{2}
      \and B.L. Jensen \inst{2}
      \and L.K. Hunt \inst{3}
      \and J. Gorosabel \inst{4}
      \and P. M{\o}ller \inst{5}
      \and J. Fynbo \inst{2,5}
      \and R. M. Kippen \inst{6}
      \and B. Thomsen \inst{7}
      \and L.F. Olsen \inst{2}
      \and L. Christensen \inst{2}
      \and M. Vestergaard \inst{8}
      \and N. Masetti \inst{9}
      \and E. Palazzi \inst{9}
      \and K. Hurley \inst{10}
      \and T. Cline \inst{11}
      \and L. Kaper \inst{13}
      \and A.O. Jaunsen \inst{14}
          }

   \offprints{M. I. Andersen}
   \mail{Michael.Andersen@oulu.fi}

   \institute{Division of Astronomy, University of Oulu
              P.O. Box 3000, FIN--90014 University of Oulu, Finland \\
              email: Michael.Andersen@oulu.fi
   \and       Astronomical Observatory, University of Copenhagen,
              Juliane Maries Vej 30, DK--2100 Copenhagen  {\O}, Denmark
   \and       Centro per l'Astronomia Infrarossa e lo Studio del Mezzo 
              Interstellare, CNR, Largo E. Fermi 5, 50125 Firenze, Italy
   \and       Danish Space Research Institute,
              Juliane Maries Vej 30, DK--2100 Copenhagen  {\O}, Denmark
   \and       European Southern Observatory, Karl--Schwarzschild--Stra{\ss}e 2,
              D--85748 Garching, Germany
   \and       University of Alabama in Huntsville and NASA/Marshall
              Space Flight Center, SD50, Huntsville, AL 35812, USA
   \and       Institute of Physics and Astronomy, University of Aarhus,
              DK--8000 {\AA}rhus C, Denmark
   \and       Astronomy Department, Ohio State University,
              140 West 18th Avenue, Columbus, OH 43210--1173, USA
   \and       Instituto Tecnologie e Studio Radiazoni Extraterresti, CNR, 
              Via Gobetti 101, 40129 Bologna, Italy
   \and       University of California, Space Science Laboratory, Berkeley, 
              CA 94720-7450, USA
   \and       NASA Goddard Space Flight Center, Greenbelt, MD 20771, USA
   \and       Sterrenkundig Instituut ``Anton Pannekoek", Kruislaan 403,
              1098 SJ Amsterdam, the Netherlands
   \and       Institute of Theoretical Astrophysics, University of Oslo,
              PB 1029, Blindern, N--0315 Oslo, Norway
             }

   \date{Received .... ; accepted ....}

   \maketitle

\begin{abstract}

We report the discovery of the gamma-ray burst \object{GRB~000131}
and its optical afterglow. The optical identification
was made with the VLT 84 hours after the burst 
following a BATSE detection and an Inter Planetary Network localization. 
\object{GRB~000131} was a bright, long-duration GRB, with an apparent precursor 
signal 62 s prior to trigger. The afterglow was detected in ESO VLT, NTT, and 
DK1.54m follow-up observations. 
Broad-band and spectroscopic observations of the spectral energy 
distribution reveals a sharp break at optical wavelengths which is 
interpreted as a Ly$\alpha$ absorption edge at 6700~{\AA}.
This places \object{GRB~000131} at a redshift of $4.500 \pm0.015$. 
The inferred isotropic energy release in gamma rays alone was
$\sim 10^{54}~{\rm erg}$ (depending on the assumed cosmology).
The rapid power-law decay of the afterglow (index $\alpha$=2.25, similar
to bursts with a prior break in the lightcurve), however, indicates
collimated outflow, which relaxes the energy requirements by a factor
of $<$ 200.
The afterglow of \object{GRB~000131} is the first to be
identified with an 8-m class telescope. 
\keywords{
cosmology: observations -- 
gamma rays: bursts -- 
ISM: dust, extinction}
\end{abstract}

%%%%%%%%%%%%%%%%%%%%%%%%%%%%%%%%%%%%%%%%%%%%%%%%%%%%%%%%%%%%%%%%%%%%%%%%%%%%%%%%

\section{Introduction}

Since the first detections of soft X-ray \cite{costa97}, 
optical \cite{paradijs97}, and radio \cite{frail97} afterglows 
to gamma-ray bursts (GRBs), 19 optical afterglows have been detected.

For the current sample of 12 GRBs with well-determined redshifts, 
the median redshift is z$_{\rm median}$ = 1.1 with a very large
root-mean-square variation of $\sim$0.8. 
While a photometric redshift of $\approx$ 5 has been proposed for
\object{GRB~980329} \cite{fruchter99}, the highest spectroscopic redshift
determination for a GRB so far is $z = 3.418$ for GRB~971214
\cite{kulkarni98}. This redshift was determined from the spectrum 
of a faint Lyman Break galaxy that was found at the position of the
optical afterglow (see also Odewahn et al., 1998). The 
highest redshift determined directly from 
absorption lines in the spectrum of an optical afterglow is 
$z = 2.0404\pm0.0008$ for \object{GRB~000301C} \cite{smette00,jensen00}. 
Several authors have
speculated about the possibility of detecting GRBs at even higher
redshifts than achieved so far. Wijers et al.~\cite*{wijers98} 
and Lamb \& Reichart~\cite*{lamb00} estimate 
that GRBs should be detectable at very 
high redshifts ($z\gtrsim 5$) based on the assumption that GRBs are 
associated with star formation and by using models for the cosmic 
star formation rate as a function of redshift. 
Blain \& Natarajan~\cite*{blain00}
propose to use the opposite strategy, namely to use the observed 
redshift distribution of GRBs to measure the (uncertain) cosmic star 
formation rate as a function of redshift.

Unfortunately, afterglows have been detected in only $\lesssim 40$ 
percent of all well-localized (SAX, RXTE, IPN) GRBs (e.g., 
Fynbo et al.~2000). It is not clear why most GRBs with 
no apparent 
optical afterglow (`dark bursts') are so abundant. Possible explanations 
are rapid decay \cite{groot98}, reddening in the immediate environment 
and host galaxy (eg., Jensen et al.~\cite*{jensen00}), intrinsic optical
faintness \cite{taylor00}, very high redshifts \cite{lamb00} or a 
combination of these effects. To shed light on dark bursts, one 
obvious strategy is to set deeper optical limits by using large 
telescopes and to determine the decay slopes and 
colours (see Fynbo et al.~2000).

In this Letter we present the first attempt to identify the afterglow
of a GRB with an 8-m class telescope. The attempt was successful in 
that the observations led to the discovery of the afterglow and the
subsequent determination of the decay slope, spectral energy
distribution and record high redshift. In the following we document the 
gamma-ray, optical, and infrared observations, and discuss the properties 
of the gamma-ray burst and afterglow. 

\section{Detection of the gamma-ray burst}

GRB 000131 was observed by Ulysses, Konus/Wind, NEAR, and
CGRO-BATSE on 2000 Jan. 31.624 UT.  After 56 hours, it was
localized via InterPlanetary Network (IPN) timing to two
alternate 55 sq. arcmin error boxes (Hurley et al., 2000a),
one of which was ruled out by the independent BATSE location \cite{kippen00}.
{From} the Ulysses, Konus/Wind, and NEAR data its 25--100 keV 
fluence was $\sim 10^{-5}$ erg/cm$^2$.

BATSE detected the event (Trigger \#7975) in a partial data gap, so the
standard catalog 
data products such as flux, fluence, and duration are not available. 
However, by analyzing other data types from BATSE the relevant parameters 
could be estimated. The results are given in Table~\ref{BATSE_data.tab}.
The peak flux is in the top 5$\%$, and the fluence in the top  
7$\%$ of all BATSE GRBs. The spectrum is well fit by a standard GRB model 
\cite{1993ApJ...413..281B} and shows a typical spectral evolution 
\cite{2000ApJS..126...19P}.

\begin{table}
\begin{center}
\caption{Spectral data from BATSE}
\scriptsize
\begin{tabular}{ l l }
\hline
Peak Flux:  &  $7.89 \pm 0.08$ ph~cm$^{-2}$~s$^{-1}$ (46--313 keV; 1.024 s) \\
Fluence:    & ($3.51 \pm 0.08$) $\times 10^{-5}$ erg~cm$^{-2}$ (26--1800 keV; 137 s) \\
Duration:   & T90 = 96.3 s (error $\sim$ 1 s) \\
            & T50 = 30.7 s (error $\sim$ 1 s) \\
Time-averaged & E$_{\rm peak} = 163 \pm 13$ keV \\
spectral      & Low-energy index $ = -1.2 \pm 0.1$ \\
parameters:   & High-energy index $ = -2.4 \pm 0.1$ \\
\hline
\label{BATSE_data.tab}
\end{tabular}
\end{center}
\end{table}

\begin{figure}
\resizebox{\hsize}{!}{\includegraphics{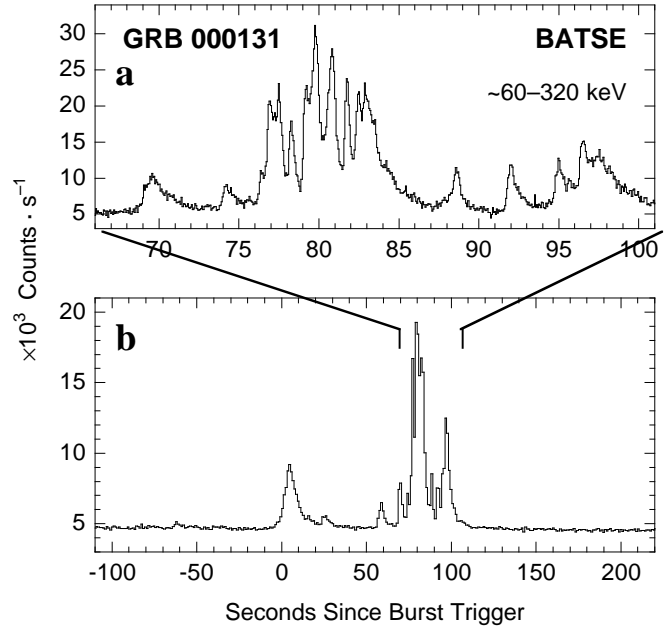}}
\put(-210,205){{\large{\bf a}}}
\put(-210,105){{\large{\bf b}}}
\caption{Gamma-ray light curve of \object{GRB 000131} recorded by the BATSE 
Large Area Detectors (LADs). {\bf b} shows continuously recorded data
with 1024 ms time resolution, {\bf a} shows an
enlargement of the most intense interval using triggered data with 64
ms time resolution. In both cases, data from the two most brightly
illuminated LADs have been summed.  Energy channels for the different
data types and detectors have been chosen to approximately match
(within a few percent) the energy range 60--320 keV.}
\label{BATSElc2.fig}
\end{figure}

The BATSE light curve is shown in Fig.~\ref{BATSElc2.fig}. 
The event consists of several spikes, most of which show a clear asymmetry,
the leading edge being the steepest.
A weak pulse lasting $\approx$~7~s is observed 62~s prior to the burst trigger.
Its arrival direction is consistent (within a 12 degree error) with that of 
the main pulse. This spike is therefore most likely part of the burst.
In this case the overall duration is about 170 s, i.e. significantly longer 
than measured by T90.
The presence of this `precursor' is reminiscent of that displayed
by GRB~991216 \cite{kippen99,hurley00a}. 
However, in the case of GRB~000131, no long-lasting
tail ($\gamma$-ray afterglow) is observed.

\section{Localization and identification of the afterglow}

The late localization of the GRB complicated the detection 
of the optical afterglow, expected to be 
significantly fainter than the limit of existing sky surveys. Furthermore, 
the (presumed power-law) decay of the afterglow would be so slow that 
observations separated by several days would have to be compared before any 
probable candidate source could be established. 

\begin{figure}
\resizebox{\hsize}{!}{\includegraphics{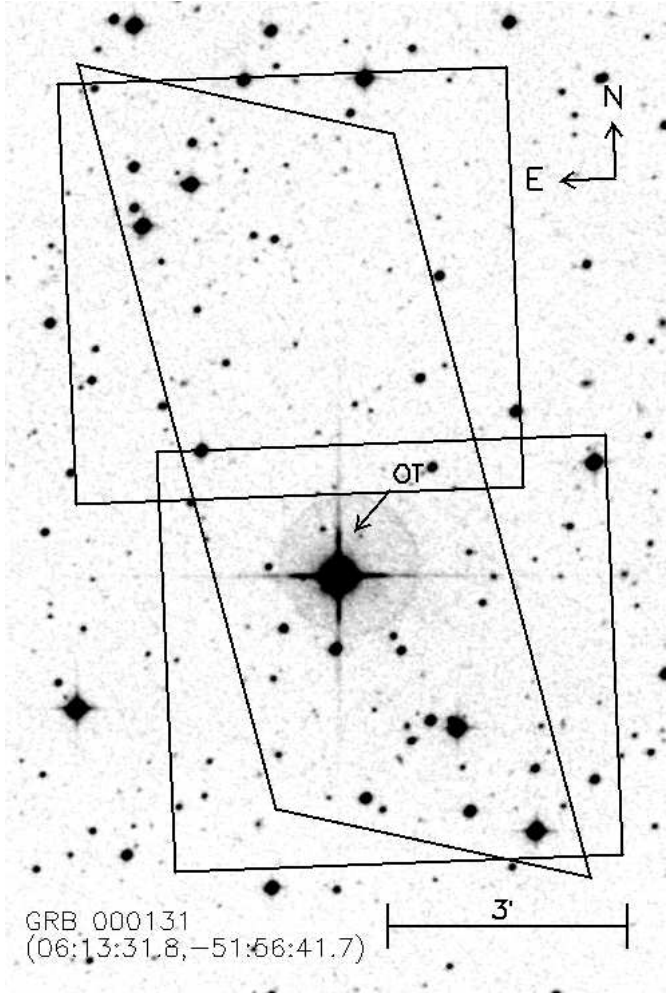}}
\caption{The rhombus shaped IPN error box with the two square FORS1 fields
superimposed. The location of the OT (not present in this Digital Sky Survey
image) is marked.}
\label{IPN_box.fig}
\end{figure}

Images of the Inter Planetary Network (IPN) error box were obtained with 
the FORS1 instrument on 
Antu (ESO VLT UT1), starting 84~h after the burst. Four 120~s exposures in
the B,V and R bands of each of two FORS1 fields, covering the error box
(see Fig.~\ref{IPN_box.fig}), were acquired under good seeing conditions
(see Table~\ref{std_mags.tab} for a log of the observations).
On the same night we acquired I--band exposures with a total integration 
time of 3600~s, using DFOSC on the Danish 1.54-m telescope on La Silla.
{From} this first set of exposures no candidate optical transient could be 
identified. 
A subsequent set of images with the same exposure times was acquired 
135~h after the burst, under less favorable seeing conditions.
Comparing these two epochs, one source located at 
R.A. = 6$^\mathrm{h}$13$^\mathrm{m}$31\fs 08, 
Dec. = $-51$\degr 56\arcmin 41\farcs 7 (J2000)
was found to have declined by about 1.1 mag in the R--band.
In the first epoch images this source was also detected in the V--band, 
but not in the B--band. Its proximity (1\farcm 1) 
to the center of the error box, in combination with a fading of about 
half a magnitude per day, which is typical for GRB afterglows after some 
days, made this object the likely afterglow of \object{GRB~000131}
\cite{pedersen00}. 
The afterglow of \object{GRB~000131} was the second to be 
identified based solely on an IPN localization \cite{hurley00a}.

\section{Follow-up observations and data analysis}

Additional images of the field were obtained on 2000 Feb. 8.1 UT. The source
was detected in a deep FORS1 R--band exposure and was found to have faded
by 1.9$\pm$0.2 mag relative to the first epoch confirming its transient nature. 
As the source, based on the V, R and I photometry, appeared to be very red, we
also acquired near-infrared (JHKs) images with the ESO New Technology 
Telescope (NTT).
A final set of R and I band images were acquired with FORS1 on 2000 Mar.~5.0 UT.
A log of our photometric observations and the standard calibrated 
magnitudes of the afterglow is given in Table~\ref{std_mags.tab}.
All image reductions were performed in the
IRAF\footnote{IRAF is the Image Analysis and Reduction Facility
made available to the astronomical community by the National Optical
Astronomy Observatories, which are operated by AURA, Inc., under
contract with the U.S. National Science Foundation.} environment.

\begin{table}
\begin{center}
\caption{Journal of our observations of \object{GRB~000131}.}
\scriptsize
\begin{tabular}{ l l @{ } l c c r r @{} l }
Telescope  & \multicolumn{2}{c}{Epoch} & Seeing     & Band &  T$_{\rm exp}$ & & ~~ mag \\
           & \multicolumn{2}{c}{(2000,UT)} & (\arcsec) &        &  (sec)  \\
\hline
VLT   & Feb&4.133   & 0.9 & B &  480 & $>$&25.85             \\
VLT   & Feb&4.135   & 0.8 & V &  480 &    &24.66 $\pm$0.10   \\ 
VLT   & Feb&4.137   & 0.7 & R &  480 &    &23.26 $\pm$0.04   \\ 
D1.54 & Feb&4.190   & 1.2 & I & 3600 &    &22.03 $\pm$0.07   \\ \hline
VLT   & Feb&6.176   & 1.3 & V &  480 & $>$&25.40             \\ 
VLT   & Feb&6.183   & 1.2 & R &  480 &    &24.35 $\pm$0.14   \\ 
D1.54 & Feb&6.170   & 1.1 & I & 3600 & $>$&23.70             \\ \hline
VLT   & Feb&8.091   & 0.8 & R &  600 &    &25.13 $\pm$0.19   \\ 
NTT   & Feb&8.046   & 0.7 & J & 1680 & $>$&22.45             \\
NTT   & Feb&8.076   & 0.7 & H & 1800 &    &22.25 $\pm$0.30   \\
NTT   & Feb&8.103   & 0.7 & Ks& 1200 &    &21.47 $\pm$0.41   \\ \hline
VLT   & Mar&5.025   & 0.9 & R &  900 & $>$&25.70             \\
VLT   & Mar&5.017   & 1.0 & I &  900 & $>$&24.85             \\ 
\hline
\end{tabular}
\label{std_mags.tab}
\end{center}
\end{table}

\subsection{Image analysis and photometry}

The afterglow was detected at low signal-to-noise in most of the images,
which made it essential to use PSF photometry to derive the magnitudes.
We used DAOPHOT-II \cite{1987PASP...99..191S} within IRAF to derive 
PSF magnitudes of all point sources in the images. 
In general, when deriving PSF magnitudes, the center and amplitude 
(magnitude) of the PSF is fitted. 
For objects close to the detection limit, fitting the centroid introduces a 
bias in the magnitude of up to one 
magnitude, because of a tendency to fit the PSF to a nearby noise spike.
If instead DAOPHOT is used on objects with accurately
known pixel coordinates, the magnitude can be derived without
re-centering the object during the PSF fit. In addition to removing the
magnitude bias, deriving the PSF magnitude without re-centering also has the 
advantage that magnitudes down to the 2-$\sigma$ level can be obtained.
We could therefore use the accurate position of the afterglow,
as determined from the first epoch images, 
to derive reliable magnitudes from the observations at later epochs. 
Limiting magnitudes were obtaind by adding a set of artificial point
sources to the images at locations of apparently blank sky, setting
the limiting magnitude at the level where DAOPHOT ALLSTAR could recover
95\% of the point sources.

\begin{figure}
\resizebox{\hsize}{!}{\includegraphics{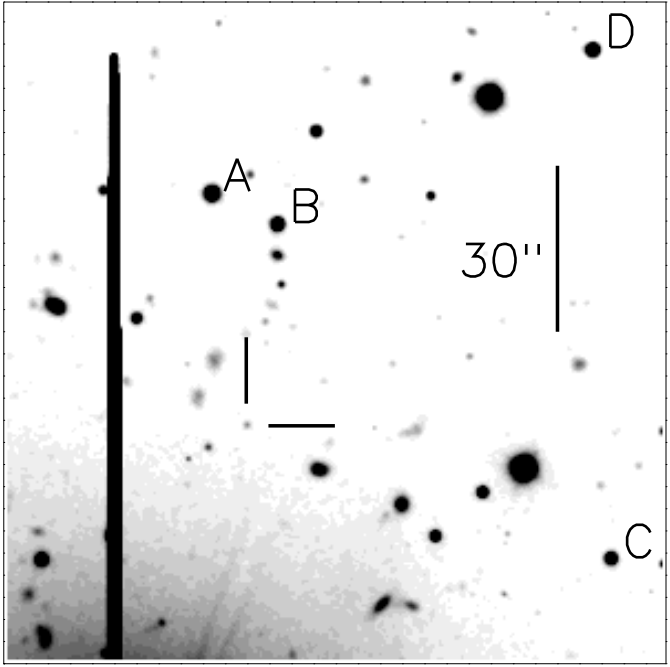}}
\caption{R--band image of the \object{GRB~000131} afterglow from Feb.~4.
The afterglow is marked with two bars and the four comparison stars
are labeled according to Table~\ref{comp_mags.tab}. North is up and
East is to the left.}
\label{discovery.fig}
\end{figure}

\begin{table}
\caption{Magnitudes of internal reference stars.}
\scriptsize
\begin{center}
\begin{tabular}{ c c c c c c c c }
Star  & B & V & R & I & J & H & K \\
\hline
A &  18.97 &  18.49 &  18.13 &  17.77 &  17.38 &  17.01 &  16.88\\
B &  21.60 &  20.07 &  18.86 &  17.48 &  16.12 &  15.53 &  15.25\\
C &  20.79 &  19.90 &  19.32 &  18.86 &  18.19 &  17.59 &  17.53\\
D &  21.15 &  19.68 &  18.65 &  17.67 &  16.54 &  15.92 &  15.73\\
\hline
\end{tabular}
\label{comp_mags.tab}
\end{center}
\end{table}

\subsection{Optical images}

  The optical observations were all acquired on nights with photometric 
sky quality. The photometry was calibrated to standard star fields 
\cite{1992AJ....104..340L} through CCD aperture photometry
relative to the comparison stars marked in Fig.~\ref{discovery.fig} and
listed in Table~\ref{comp_mags.tab}.
The errors in the magnitudes of the reference stars are dominated
by the determination of the photometric zero points.
Extinction correction was applied using observatory standard extinction values.
Because the standard star exposures were acquired at approximately the 
same airmass as the object exposures, corrections were in all cases 
smaller than the individual photometric error of the afterglow
magnitude. From the errors in the color transformation, it is estimated
that the photometric zero point errors are smaller than 0.02 mag. 
In the R and I images from Mar.~5, there is no detection of a source within
2\arcsec\ of the afterglow location. This implies an upper limit to the host
galaxy magnitude of R = 25.7 and I = 24.8.

\subsection{Near-infrared images}

Near-infrared (IR) J, H, and Ks--band images were obtained at La Silla 
with the NTT and SOFI on 2000 Feb.~8.1 UT. SOFI is equipped with a Hawaii 
1024$\times$1024 pixel HgCdTe 
detector, and we used a plate scale of 0\farcs29 which gives a field-of-view
of roughly 4\farcm9$\times$4\farcm9.
Each image comprises several tens of elementary co-added frames acquired
by randomly dithering the telescope several arcsec once a minute.

The frames were reduced by first subtracting a mean sky,
obtained from frames acquired just before and after the source frame.
Before using the frames for sky subtraction, stars were eliminated with
an automatic star finder coupled with a background interpolation
algorithm.
Then, a differential dome flat-field correction was applied, and the
frames were registered to fractional pixels and combined.
We calibrated the photometry with standard stars taken from
Persson et al.~\cite*{1998AJ....116.2475P}, acquired directly before and
after the source observations.
The standards were placed in five different positions on the detector,
and reduced in the same way as the source frames.
Standard star and source photometry was corrected for atmospheric
extinction using the mean ESO La Silla extinction coefficients given
in Engels et al.~\cite*{1981A&AS...45....5E}.
Formal photometric accuracy, as judged from the standard deviation
of the standard star observations, is about 0.03 mag.

At the time of observations, about 8 days after
the IPN detection, the afterglow was very faint and thus only marginally
detected in H (3 $\sigma$) and K (2.5 $\sigma$).
In J there is apparently a source,
but it is displaced about 1 pixel from the location of the afterglow. 
A PSF magnitude of this source could only be derived if re-centering during 
the PSF fit was allowed. To maintain consistency in our magnitudes, we have 
therefore chosen to disregard this apparent source and only give the proper
limiting magnitude. The H and K magnitudes are both very close to the 
detection limit. It was verified that magnitudes at this level can be reliably 
derived by performing photometry of a large number of artificially added stars
of the same magnitude.

\subsection{Spectroscopy}

Starting 2000 Feb.~8.09 UT, we obtained spectroscopic observations, using FORS1
configured with a 300 lines/mm grism, a cut-on order sorter filter and 
a 0\farcs7 slit. The wavelength
range covered was 3800~{\AA} -- 8200~{\AA}, with a resolution of 9~{\AA} 
and with order overlap from about 7600~{\AA}.
The total integration time was 10800~s, divided into 6 exposures of 1800~s.
During the spectroscopic observations, the seeing ranged from 0\farcs55 to
0\farcs9. The combined spectrum was flux calibrated relative to the flux 
standard \object{LTT3218}.
The continuum is not well detected at visible wavelengths but is 
weakly detected in the red part of the spectrum, with a signal-to-noise 
around 1 per spectral bin of 2.5{\AA} in regions that are not dominated by 
sky lines. The spectral slope is consistent with the broad-band photometry. 
The spectrum does not show signs of statistically significant emission lines,
as could be expected from a host galaxy.

\section{Discussion}

The temporal and spectral behavior of optical gamma-ray burst afterglows
are in general observed to follow a power law, such that
$F_\nu (t,\nu) \propto t^{-\alpha} \nu^{-\beta}$ ($\alpha,\beta>0$), where $t$
is the time elapsed since the GRB event. The fireball model provides
the theoretical framework for this behavior and also gives the relation 
between the temporal slope $\alpha$ and the spectral slope $\beta$ in 
various regimes
\cite{1998ApJ...497L..17S,1999PhR...314..575P,1998ApJ...499..301M}.
If the afterglow is collimated (a jet), the temporal slope is modified 
and may change during the transition from the observer being inside the 
jet cone to being outside the jet cone, caused by the slowing-down of the 
jet \cite{rhoads99}.  This results in a break in the light curve, such as 
those observed in 
\object{GRB~980519} \cite{jaunsen00},
\object{GRB~990123}, 
\object{GRB~990510} \cite{stanek99,harrison99},
\object{GRB~990705} \cite{masetti00a},
\object{GRB~991216} \cite{halpern00}, and
\object{GRB~000131C} \cite{jensen00,masetti00b}.
In the following, we will base our discussion on the assumption that
the afterglow can be described within the framework of the fireball model.

\subsection{The light curve of the afterglow }

\begin{figure}
\resizebox{\hsize}{!}{\includegraphics{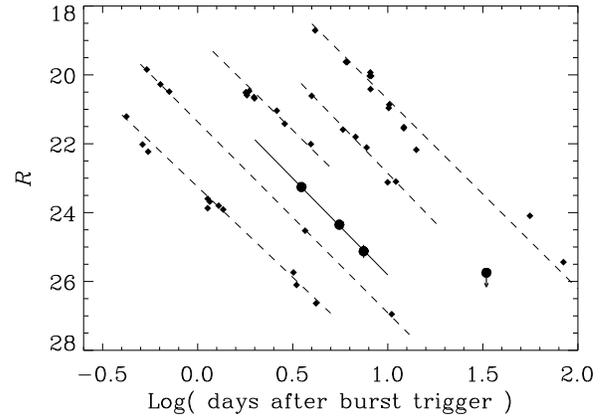}}
\caption{R--band light curves of GRB afterglows with steep late-time decay.
The solid line is a power-law fit to the GRB 000131 data (filled circles).
The non-detection on March 5 is marked as an upper limit at $\log{t}$=1.52.
Other afterglows are (from left to right) \object{GRB~980326}, 
\object{GRB~980519}, \object{GRB~990510}, \object{GRB~000301C} and
\object{GRB~991208} (which is shifted 0.4 in log(days)). Pre-break data
points are omitted. }
\label{lightcurve.fig}
\end{figure}

The three epochs of R--band photometry can be fitted by a power-law decline
with $\alpha = 2.25 \pm 0.19$ and a $\chi^2 = 0.06$, as shown in
Fig.~\ref{lightcurve.fig}.  
A power-law decline with this slope is typical for post-break evolution of 
GRB afterglow emission.  A value of $2.0<\alpha< 2.5$ has been observed
in several other afterglows, namely
\object{GRB~980326} \cite{groot98},
\object{GRB~980519} \cite{jaunsen00},
\object{GRB~990510} \cite{stanek99,harrison99,holland00},
\object{GRB~991208} \cite{hurley99,castro-tirado00}, and
\object{GRB~000301C} (e.g., Jensen et al.~2000). 
The light curve of \object{GRB~000131} is plotted together with the 
light curves of these afterglows in Fig.~\ref{lightcurve.fig}.
The main characteristics of these systems are summarized in 
Table~\ref{alpha_and_betas.tab}. In all cases the favoured
interpretation of such a steep slope is post-break decay of a collimated
outflow. The light-curve breaks predicted in this scenario have indeed 
been observed (see Holland et al.~2000) and are believed to have occurred
prior to the first observations in the other systems with $\alpha > 2$.
The simplest interpretation of the \object{GRB~000131} afterglow
light curve is therefore that of a collimated outflow seen after 
the jet has slowed down.

\begin{table}
\begin{center}
\caption{GRB afterglows with rapidly decaying light curves.}
\begin{tabular}{ l c c c c }
~GRB & $\alpha$ & $\beta$ & t$_\mathrm{break}$ & R($t$=4~d) \\
\hline
980326  & 2.10 $\pm$0.13 & 0.66 $\pm$0.70 &                 & 26.4 \\
980519  & 2.22 $\pm$0.04 & 0.80 $\pm$0.08 & 0.55 $\pm$0.17  & 24.7 \\
990510  & 2.18 $\pm$0.05 & 0.61 $\pm$0.12 & 0.80 $\pm$0.35  & 22.1 \\
991208  & 2.20 $\pm$0.20 & 0.75 $\pm$0.03 &                 & 20.2 \\
000131  & 2.25 $\pm$0.19 &                &                 & 23.6 \\
000301C & 2.29 $\pm$1.00 & 0.55 $\pm$0.04 & 4.39 $\pm$1.52  & 20.6 \\
\hline
\end{tabular}
\label{alpha_and_betas.tab}
\end{center}
\end{table}

\subsection{Spectral energy distribution}

\begin{figure}
\resizebox{\hsize}{!}{\includegraphics{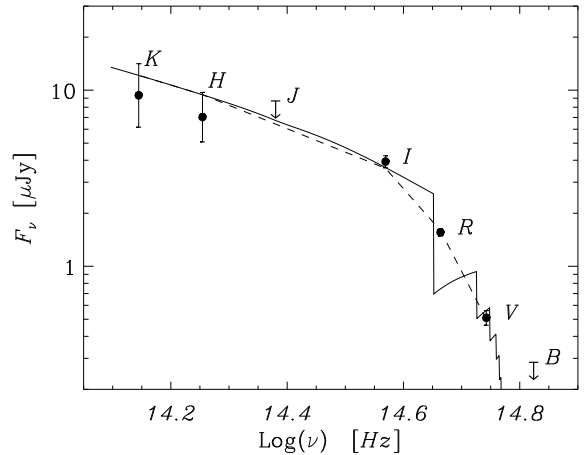}}
\caption{The spectral energy distribution, as derived from broad band
photometry. The errors of the H and K fluxes include the formal error
from the extrapolation of the light curve back to $t$=3.5 days.
A fit by a power law spectrum with Lyman forest absorption and
SMC reddening is shown as a dashed line. This yields A$\mathrm{_V}$ = 0.18,
when an intrinsic spectral slope, $\beta$ = 0.70, and a redshift of 4.5 is
assumed. The solid line shows the corresponding
spectrum with its Lyman absorption edges.}
\label{sed.fig}
\end{figure}

The optical afterglow was detected in VRIHK, but at different epochs.
By assuming achromatic evolution of the afterglow (which is
consistent with other afterglows), following the
$\alpha$ = 2.25 power-law decline, the spectral energy distribution can 
be calculated for a given epoch. In Fig.~\ref{sed.fig} we plot the derived 
spectral energy distribution at 3.50 days after the burst trigger, i.e., at 
the time of the first VRI observations. Also given are the upper limits derived 
from our B and J observations. The spectral energy distribution has been 
corrected for Galactic reddening, using E$_{\mathrm{(B-V)}}$=0.056 
\cite{1998ApJ...500..525S}. 

For steep-decay afterglows, like those shown in Fig.~\ref{lightcurve.fig},
the large values of 
$\alpha$ and the relatively small values of $\beta$ favour a scenario 
involving a sideways expanding jet. In this case spectral slopes between 
$(\alpha -1)/2$ and $\alpha /2$ are expected in the fireball model,
depending on the value of the cooling frequency. With $\alpha = 2.25$,
this implies $0.63 < \beta < 1.13$, with a preference for the low
value (cf.~Table~\ref{alpha_and_betas.tab}). 

The spectral energy distribution shown in Fig.~\ref{sed.fig} 
does not resemble a power-law with a single index $\beta$. This
is not an artifact of extrapolating the H and K band data points from 
Feb.~8 to Feb.~4, as a power-law fit to the V, R, and I data 
points only results in a $\chi^2 = 7.1$ and $\beta = 4.90$, which is 
much steeper than the value of about 1 typically observed in GRB afterglows 
(see Table~\ref{alpha_and_betas.tab}). Moreover, for the H and K fluxes 
to be in accordance with this fit, an unphysical value of $\alpha$ of 
about 7.0 in the near-IR is required.  Thus, a power-law spectral energy 
distribution is ruled out. 

We have explored whether the strongly curved shape of the spectral
energy distribution can be explained by reddening in the host galaxy
(as in eg.\ \object{GRB~000301C}, Jensen et al.~2000, or in
\object{GRB 971214}, Dal Fiume et al.~2000). We find that
no physically plausible reddening laws can transform a power-law
spectrum into the observed shape.

The most likely interpretation of the spectral break is therefore 
the onset of Lyman forest blanketing, hence implying that 
\object{GRB~000131} was at a high redshift (z$\gtrsim$4). To 
examine this interpretation further we have fitted the spectral energy 
distribution with power-laws modified by the effects of Lyman forest 
blanketing and internal reddening in the host galaxy. This technique was 
first used by Fruchter \cite*{fruchter99} for \object{GRB~980329} and discussed 
in detail by Reichart \cite*{reichart00}.
For the reddening we use the SMC extinction law assumed to best
represent a chemically less-evolved environment at redshifts of 4 to 5.
To model the effects of Lyman forest blanketing we follow M\o ller and
Jakobsen~\cite*{moller90}: 
for each pair of values of $\beta$ and $z$, the visual rest-frame absorption 
A$_{\mathrm{V}}$ and the $\chi^2$ of the fit is obtained. We find that all 
values of $\beta$ allowed by the fireball model are possible, with a 
preference for low values. The corresponding range of internal reddening 
is 0.11 $<$ A$_{\mathrm{V}}$ $<$ 0.20, with a preference for high values.
For values of $\beta$ around 0.65, the possible 
range (2 $\sigma$) of redshifts is $4.38 < z < 4.70$, while the redshift
is $4.55 < z < 4.69$ for $\beta$ = 1.0.
This result is not affected by the choice of reddening law.

\begin{figure}
\resizebox{\hsize}{!}{\includegraphics{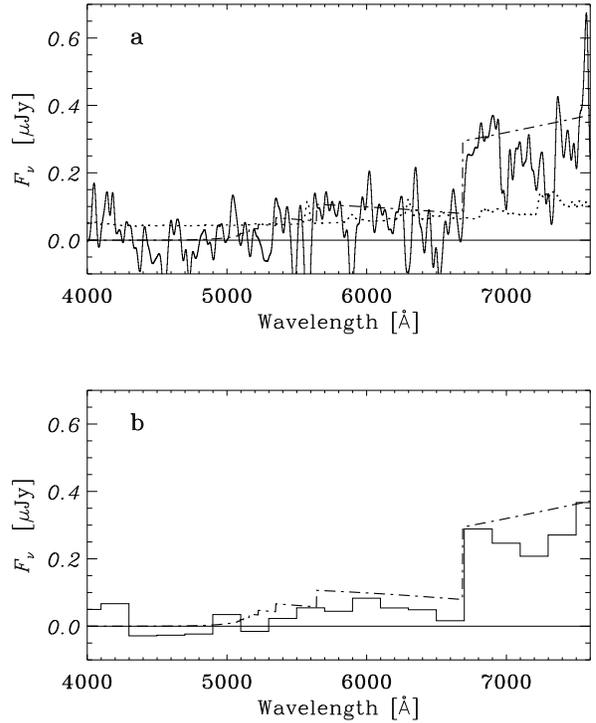}}
\caption{The spectrum of the \object{GRB~000131} afterglow:
{\bf a} smoothed to a resolution of 30~{\AA}, {\bf b} rebinned to a resolution
of 200~{\AA}. The dotted curve shows the noise per bin, while the
dot-dashed line is the model spectrum shown in Fig.~\ref{sed.fig}.}
\label{spectrum.fig}
\end{figure}
% HERE IS A PROBLEM

\subsection{The spectroscopic redshift of \object{GRB~000131}}

Our analysis of the photometric observations implies the presence of a
Ly$\alpha$ absorption edge in the range 6500~{\AA} to 6900~{\AA}. 
This implication is confirmed by the spectroscopic observations.
After smoothing the FORS1 spectrum to a resolution of 30~{\AA} a clear 
indication of such an absorption edge at 6700~{\AA} is revealed.
The recorded spectrum was very faint, at the level of less than 2\% of the 
continuum of the night sky, which is dominated by emission lines through
the red part of the spectrum.
The noise in the spectrum is therefore dominated by uncertainty in the sky 
subtraction. An improved representation of the continuum was obtained 
by rebinning the spectrum to a resolution of 200~{\AA}, 
omitting spectral bins coincident with sky lines. The smoothed
and the rebinned spectra are shown in Fig.~\ref{spectrum.fig} 
together with a model of the spectrum, assuming a redshift of 4.5, 
$\beta$ = 0.70 and a reddening, A$_{\mathrm{V}} = 0.18$, as derived 
from a fit to the photometry (see Fig.~\ref{sed.fig}). 

The model spectrum is normalized to the R-magnitude of the afterglow,
obtained immediately before the spectroscopic observations began,
and corrected for slit losses of 35\%, as would result from using a
0\farcs7 slit in a seeing of about 0\farcs7. The model spectrum
is seen to be in very good agreement with the binned spectrum for
wavelengths below 7200~{\AA}. The average observed flux in the Lyman
forest region (between Ly$\alpha$ and Ly$\beta$) is $55 \pm15$~nJy,
which is in reasonable agreement with the level of 79~nJy predicted
by the model. Beyond 7200~{\AA} the spectrum is dominated by an
atmospheric band and strong sky lines, which are not well resolved at a
spectral resolution of 9~{\AA}. This effectively renders the spectrum useless
in this spectral region. 

\begin{figure}
\resizebox{\hsize}{!}{\includegraphics{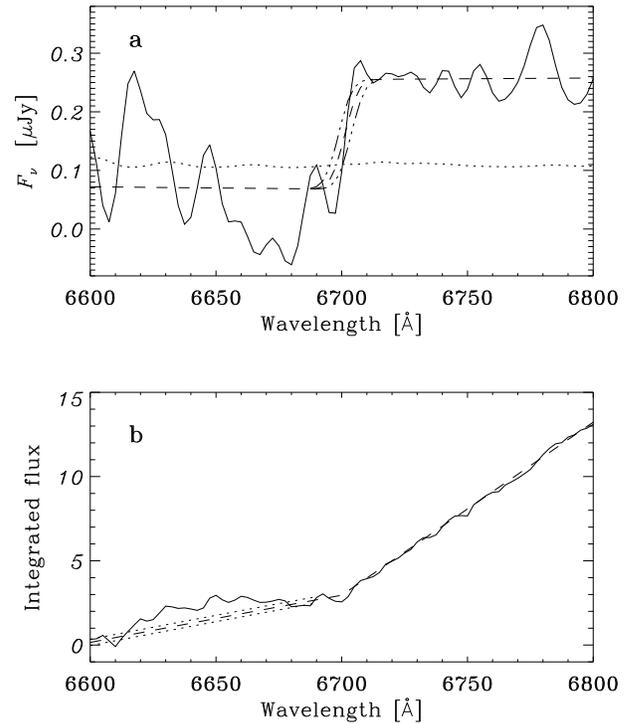}}
\caption{{\bf a} A region of the spectrum, centered on 
the Ly$\alpha$ absorption edge and smoothed to a resolution of 13~{\AA}.
The dotted line gives the noise per bin, while the dashed curve is a model
absorption edge spectrum corresponding to a redshift of 4.511, and redshifts
of 4.509 and 4.513 (dot-dashed curves). {\bf b} The integrated
flux of the full resolution spectrum and the model absorption edge spectra
shown in {\bf a}. The integrated flux is normalized to a common 
level at 6750~{\AA} \label{lyman_edge.fig}.}
\end{figure}

As the spectral region from 6590~{\AA} to 6810~{\AA} is essentially  free 
from sky lines, the location of the absorption edge is well
defined (see Fig.~\ref{lyman_edge.fig}). We measure its location to be 
at $6701 \pm2.5$~{\AA} from which a redshift of $4.511 \pm0.002$ is inferred.
The reality of the absorption edge and the interpretation that it is due 
to Ly$\alpha$ is most convincingly seen by integrating the flux of the 
spectrum and the model, as shown in Fig.~\ref{lyman_edge.fig}. The integrated 
spectrum is very smooth and follows the model nicely on the red side 
of the edge, while there are significant deviations on the blue side 
of the edge. Comparing 100~{\AA} intervals on the blue and red side of
the absorption edge, the standard deviation of the spectrum 
is found to be 65\% larger on the blue side, consistent with the 
interpretation that the blue side is located in the Lyman forest.
Hence, the VLT spectrum is in excellent agreement with the photometric
observations and provides independent evidence that the redshift 
of GRB~000131 is $z\approx4.5$.
   
In the VLT spectrum of \object{GRB~000301C}, Jensen et al. \cite*{jensen00}
detected a strong damped Ly$\alpha$ absorption line at the redshift of the 
GRB. If a similar damped Ly$\alpha$ absorption line is 
present in the spectrum of \object{GRB~000131} the redshift as determined 
from the spectral break will be slightly overestimated since the red wing of 
the damping profile will move the spectral break 10--20~\AA\ (depending on the
\ion{H}{i} column density) towards the red. Therefore, the inferred
redshift from the spectral break depends on the assumed \ion{H}{i}
column density of the GRB self absorption. With the signal-to-noise of the
spectrum, it is not possible to constrain the \ion{H}{i} column density.
An indirect indication that \object{GRB~000131} had
significant self absorption is obtained from the estimate of A$\mathrm{_V}$.
Assuming an SMC extinction law and minimum dust-to-gas ratio
\cite{1992ApJ...395..130P} and the maximum value of A$\mathrm{_V}$,
as derived from the fit of the spectral energy distribution, 
an \ion{H}{i} column density of up to $10^{22}$~cm$^{-2}$ is estimated.
An estimate of the likely error of the redshift may therefore be obtained
by fitting a \ion{H}{i} line corresponding to this column density to the 
spectrum. We derive a redshift of 4.490~$\pm$0.002 for this column density,
which allows us to conservatively 
conclude that the redshift of \object{GRB~000131} is $z = 4.500\pm0.015$.

\section{Conclusions}

Based on the temporal and positional coincidence of the transient object with
\object{GRB~000131}, and the fact that it was located at a cosmological 
distance, we conclude that this transient object was the optical afterglow
of \object{GRB~000131}.
Assuming a Hubble constant of 65 km~s$^{-1}$~Mpc$^{-1}$,
$\Omega_0=0.3$ and $\Omega_\Lambda=0.7$, a redshift of 4.50 implies a distance
modulus of 48.24. Hence, the total energy release assuming isotropic
emission is $\approx$1.1$\times$10$^{54}$ erg. This is only a factor of
$\sim$3 smaller than the isotropic equivalent energy for the most
energetic event so far, \object{GRB~990123} 
\cite{andersen99,1999Natur.398..389K}. However, the combination of
the decay slope $\alpha$ and the constraints on the spectral slope
$\beta$ indicate that the afterglow was due to a collimated jet.
The upper limit on the break epoch, t$_{\rm break}$ $<$ 3.5 days, implies 
a lower limit on the jet opening angle of 
$\theta >$ 7$^{\circ}$ $n^{1/8}$ $>$ 7$^{\circ}$, 
where $n$ is the density of the ambient medium (in units of cm$^{-3}$), 
which is larger than 1 in star-forming regions. From the lower limit on the 
opening angle we infer a lower limit on the released energy of 
$\sim5\times10^{51}$ erg.

This work shows that 8-m class telescopes may be used successfully
to detect faint R~$>23$ optical afterglows even in fields with very
bright stars. This is necessary in order to resolve why $\gtrsim60\%$
of all attempts to detect optical afterglows of GRBs are currently
unsuccessful \cite{fynbo00}. 
The apparent brightness of the \object{GRB~000131} afterglow was similar
to that of \object{GRB~990123}. \object{GRB~000131} does 
therefore provide the first observational evidence that it is possible
to obtain high resolution optical spectra of GRBs at very high redshift,
if the afterglow is identified at an early time. High redshift GRBs will 
therefore no doubt prove to be an extremely valuable tool for not only
the understanding of the GRB environment but also for the study of the 
Lyman forest and of cosmology.

\section*{Acknowledgements}
MA acknowledges the Astrophysics group of the Physics dept. of 
University of Oulu for support of his work. This work was supported by the 
Danish Natural Science Research Council (SNF).
Part of the data presented here were obtained as part of an 
ESO Service Mode programme. The allocation of observing time by IJAF
at the Danish 1.54~m telescope on La Silla is acknowledged.
KH is grateful for support under JPL Contract 958056 for Ulysses operations,
and to NASA Grant NAG 5 9503 for NEAR operations.
We are grateful to the NEAR XGRS team and the Konus/Wind team for the use of 
their data in localizing this burst.
MV gratefully acknowledges financial support
from the Columbus Fellowship at The Ohio State University.

%\bibliography{aamnem99,astro}
% Have included the .bbl here.

\end{document}